\documentclass[useAMS,usenatbib]{mn2e}

\usepackage{verbatim,graphicx,epsfig,dcolumn}
\newcolumntype{.}{D{.}{.}{4}}
\newcolumntype{,}{D{.}{.}{2}}
\newcolumntype{;}{D{.}{.}{1}}
\newcommand{\nodata}{$\cdot\cdot\cdot$}
\newcommand{\lesssim}{{\lower-1.2pt\vbox{\hbox{\rlap{$<$}\lower5pt\vbox{\hbox{$\sim$}}}}}}
\newcommand{\gtrsim}{{\lower-1.2pt\vbox{\hbox{\rlap{$>$}\lower5pt\vbox{\hbox{$\sim$}}}}}}

\title[RGB and HB mass loss in $\omega$ Cen]{Empirical determination of the integrated red giant and horizontal branch stellar mass loss in $\omega$ Centauri}
\author[I. McDonald, C. I. Johnson \& A. A. Zijlstra]{I.~McDonald$^{1}$\thanks{E-mail: mcdonald@jb.man.ac.uk}, C.~I.~Johnson$^{2,3}$, A.~A.~Zijlstra$^{1}$\\
$^{1}$Jodrell Bank Centre for Astrophysics, Alan Turing Building, Manchester, M13 9PL, UK\\
$^{2}$Department of Physics and Astronomy, UCLA, 430 Portola Plaza, Box 951547, Los Angeles, CA 90095-1547, USA\\
$^{3}$National Science Foundation Astronomy and Astrophysics Postdoctoral Fellow}

\begin{document}

\date{Accepted 9999 December 32. Received 9999 December 32; in original form 9999 December 32}

\pagerange{\pageref{firstpage}--\pageref{lastpage}} \pubyear{9999}

\maketitle

\label{firstpage}

\begin{abstract}
We herein determine the average integrated mass loss from stars belonging to the dominant metal-poor population ([Fe/H] $\sim$ --1.7) of the Galactic globular cluster $\omega$ Centauri (NGC 5139) during their red giant and horizontal branch evolution. Masses are empirically calculated from spectroscopic measurements of surface gravity and  photometric measurements of temperature and luminosity. Systematic uncertainties prevent an absolute measurement of masses at a phase of evolution. However, the relative masses of early asymptotic giant branch stars and central red giant branch stars can be measured, and used to derive the mass loss between these two phases. This can then be used as a physical check of models of horizontal branch (HB) stars. For $\omega$ Cen, the average difference is found to be 26 $\pm$ 4 \%. Assuming initial and final masses of 0.83 and 0.53 M$_\odot$, we determine that 0.21 $\pm$ 0.03 M$_\odot$ is lost on the RGB and 0.09 $\pm$ $\sim$0.05 M$_\odot$ is lost on the AGB. The implied HB stellar mass of 0.62 $\pm$ 0.04 M$_\odot$ is commensurate with literature determinations of the masses of the cluster's HB stars. The accuracy of this measurement can be improved through better selection of stars and spectral coverage, and applied to other clusters where horizontal branch models do not currently agree.
\end{abstract}



\section{Introduction}
\label{IntroSect}

Though all stars lose mass, the vast majority experience their most significant mass loss on the red and asymptotic giant branches (RGB/AGB). The amount of mass lost determines both the mass and type of the stellar remnant. It can profoundly alter not only the post-RGB track of a star's evolution in a Hertzsprung--Russell diagram (HRD), but also the mass, chemical state, composition and mineralogy of the material returned to the interstellar medium. The processes that govern this mass loss are relatively-well understood (e.g., see review \citealt{Willson00}), and can be broadly divided into two categories. Mass loss from a star's hot ($\sim$10\,000 K) chromosphere ejects mostly atomic gas at low mass-loss rates ($\lesssim 10^{-7}$ M$_\odot$ yr$^{-1}$; \citealt{DSS09}) over most of the star's life. Conversely, pulsation-enhanced mass loss driven by a cool ($\lesssim$2000 K), dusty wind can reach rates of $>$10$^{-5}$ M$_\odot$ yr$^{-1}$ \citep{vLGdK+99}, but for a much shorter period of time.

Instantaneous mass-loss rates have been derived for a large number of stars, but they are difficult to measure with any absolute accuracy and can vary substantially on evolutionarily-short timescales. Determination of the integrated mass-loss rate (i.e.\ the difference in mass) between two evolutionary phases is necessary if one is to probe how a star with a given set of parameters will evolve. 

Stellar clusters provide excellent testbeds in which to probe stellar evolution, containing one (or sometimes a few) distinct populations at a known age, distance and metallicity, which have giant stars at the same initial mass. Increasingly-precise stellar evolution models (e.g.\ \citealt{MGB+08,DCJ+08}) and white dwarf observations (e.g.\ \citealt{MKZ+04}) can determine the initial and final masses of present giants to within a few percent. In intermediate stages, however, there are considerable systematic differences in the modelled stellar masses.



This is perhaps most notable for horizontal branch (HB) stars. In clusters with blue HBs, stellar masses are relatively easy to determine via HB modelling and RR Lyrae pulsations. If the masses of HB stars are $\gtrsim$0.65 M$_\odot$ (depending on the HB model and the cluster metallicity), they will be too cool to become RR Lyrae stars and modelling becomes less precise due to the weaker dependence of the predicted $T_{\rm eff}$ of red HB stars on their mass. In some cases, different models can disagree on HB masses by as much as $\sim$25\%, depending on the model chosen \citep{CGBC96,Catelan09,MBvL+11}. This causes severe problems when one tries to determine the total mass lost on the RGB, whether RGB or AGB mass loss is dominant, and the detailed evolutionary stages where mass loss is important.

In this paper, we use a simple method to directly determine the fractional mass difference between RGB and early-AGB stars, which by definition gives the time-integrated mass loss between these two phases. We apply this method to the metal-poor ([Fe/H] $\approx$ --1.7) population of $\omega$ Centauri (NGC 5139): a well-studied globular cluster with a blue HB and well-determined HB star masses. This allows us to empirically confirm the accuracy of the HB models and investigate the precision with which we can determine the time-integrated mass loss.

\section{Method}
\label{MethodSect}

If one knows the distance and reddening to a star, one can determine its mass simply from three observable parameters: the gravity, $\log\,g$, determined from high-resolution spectroscopy; the bolometric luminosity, $L$, derived from the integrated spectral energy distribution (SED) and cluster distance; and the effective surface temperature, $T_{\rm eff}$, derived from spectroscopy, photometric colours or SEDs. Using these, the stellar radius can be simply derived using the Stefan--Boltzmann Law, and Newtonian gravity as follows:
\begin{eqnarray}
R^2 &=& \frac{L}{4 \pi \sigma T_{\rm eff}^4} = \frac{G M}{g}, \nonumber \\
{\rm hence:} \nonumber \\
M &=& \frac{gL}{4 \pi \sigma T_{\rm eff}^4 G} .
\label{MEqn}
\end{eqnarray}
Given the difficulty in determining these parameters accurately (particularly $\log\,g$), it is not normally possible to usefully constrain the masses of individual stars. Systematic errors mean that absolute masses for groups of stars are also usually too imprecise to be useful. By applying a statistical approach to a large number of systematically-surveyed stars, however, one can gain a useful measure of the \emph{difference} in mass between two populations of stars within the same cluster that is theoretically model-independent. In practice, one must accept some dependency on the stellar atmosphere models used to derive the observables ($L$, $T_{\rm eff}$ and log\,$g$), but these can mostly be circumvented by choosing stars with a similar temperature range where systematic effects are small.

In recent years, large-scale studies of stars in globular clusters have meant many stars have had their fundamental parameters accurately determined. The cluster $\omega$ Cen has had photometric temperatures and luminosities, and spectroscopic metallicities determined for a large number of its stars (\citealt{MvLD+09}, hereafter M+09; and \citealt{JP10}, hereafter JP10, respectively; see also \citealt{vLvLS+07}; \citealt{MMP+11}). We present here the spectroscopic gravity determinations from the original spectra of JP10 and use these to measure the relative masses of RGB and AGB stars.

\section{Sample section \& data reduction}
\label{DataSect}

\begin{center}
\begin{table}
\caption{Spectroscopic and photometric parameters for our target stars. A complete table is available online.}
\label{ParamTable}
\begin{tabular}{lcccc}
    \hline \hline
LEID	& $T_{\rm phot}$& $L_{\rm phot}$	& [Fe/H]	& log\,$g$	\\
\ 	& (K)			& (L$_\odot$)	& (dex)	& (cm s$^{-2}$)	\\
    \hline
\multicolumn{5}{c}{RGB}\\
    \hline
16019	& 4803		& 206.8	& --1.74		& 1.64		\\ 
16027	& 4850		& 166.0	& --1.86		& 1.63		\\ 
\nodata& \nodata		& \nodata	& \nodata		& \nodata		\\ 
    \hline
\multicolumn{5}{c}{AGB}\\
    \hline
19022	& 4963		& 232.2	& --1.80		& 1.51		\\ 
24040	& 4907		& 221.3	& --1.75		& 1.47		\\ 
\nodata& \nodata		& \nodata	& \nodata		& \nodata		\\ 
    \hline
\end{tabular}
\end{table}
\end{center}

\begin{figure}
\centerline{\includegraphics[height=0.5\textwidth,angle=-90]{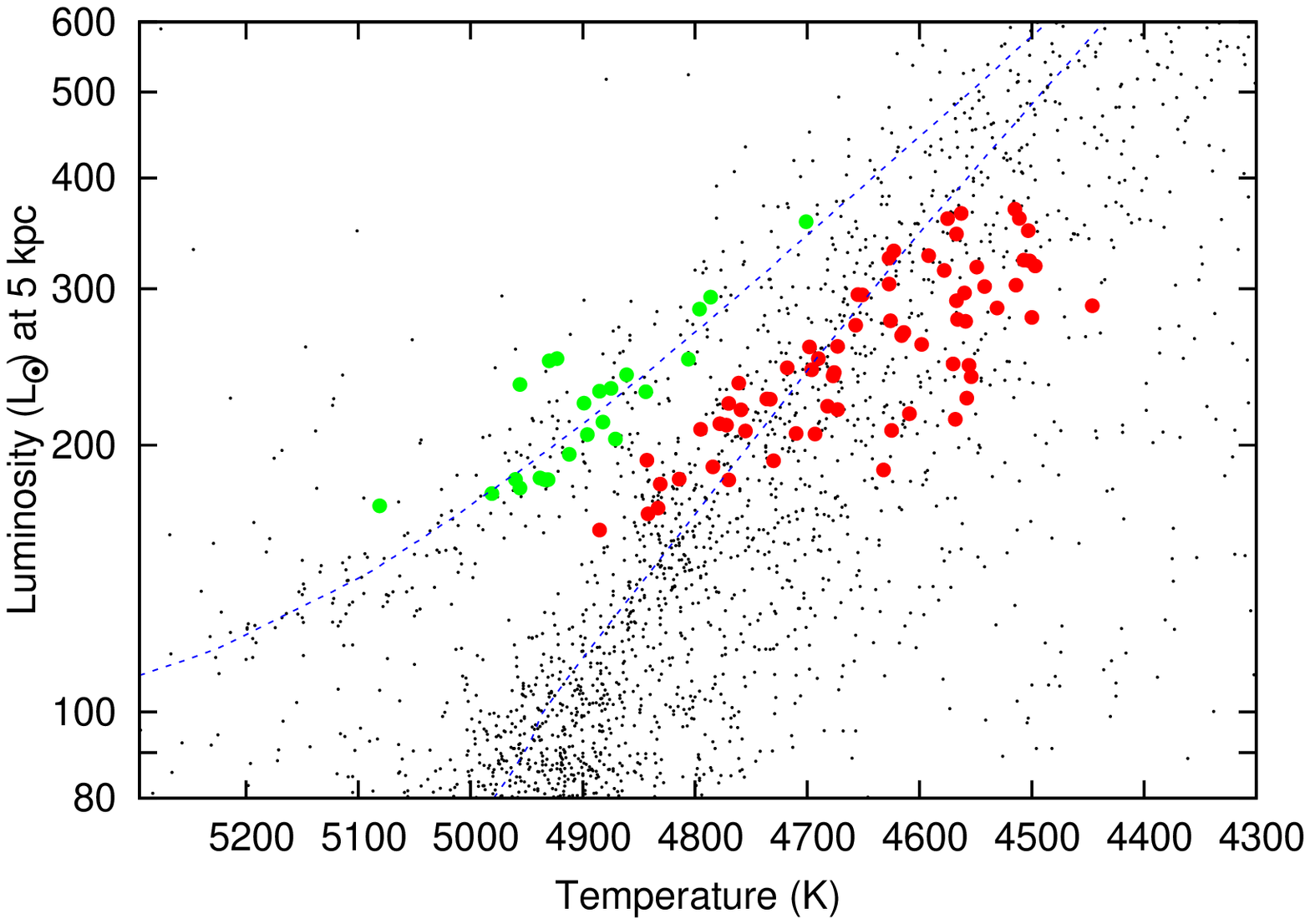}}
\centerline{\includegraphics[height=0.5\textwidth,angle=-90]{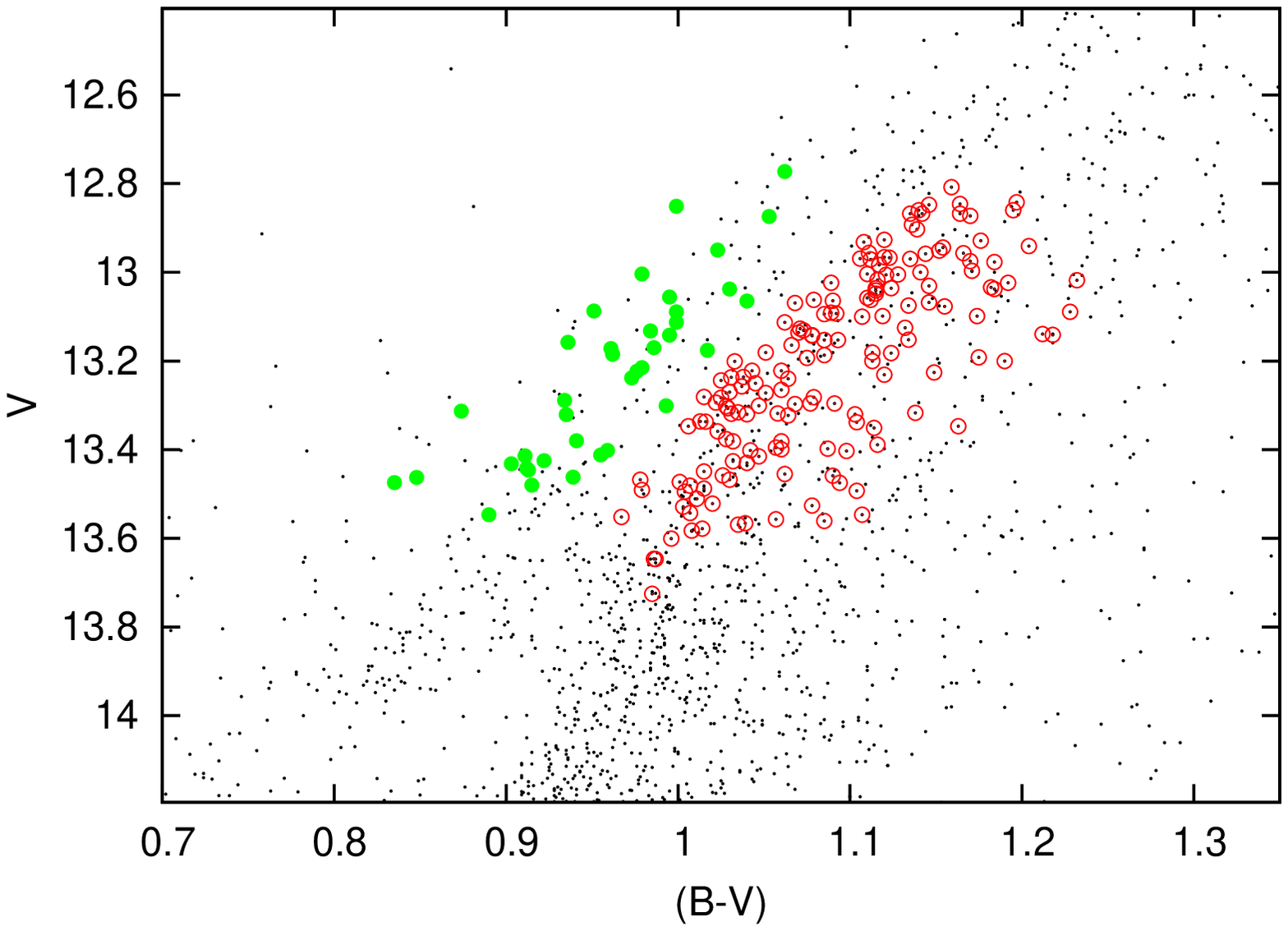}}
\caption{Hertzsprung--Russell and colour-magnitude diagrams showing the central giant branches stars in $\omega$ Cen. The larger red and green circles show our RGB and eAGB targets, respectively. An isochrone and AGB model from \protect\citet{DCJ+08} is included, set at 12 Gyr, [Fe/H] = --1.7, [$\alpha$/Fe] = +0.4.}
\label{HRDFig}
\end{figure}

An essential requirement is to accurately differentiate RGB from AGB stars. The most reliable way to do so is photometrically, on the early-AGB (eAGB), where the two branches separate on Hertzsprung--Russell and colour--magnitude diagrams (HRD, CMD). Our selection is based on cuts applied to the HRD of M+09 and the CMD of \citet{BPB+09}: stars common to M+09 and JP10 are selected if they fall into the designated regions on both the HRD and CMD (Figure \ref{HRDFig}).

A well-known spread exists in the elemental abundances and metallicity of $\omega$ Cen's stars (e.g. JP10). To minimise the effects of this spread, and due to the difficulty in identifying AGB stars in the metal-rich populations in the cluster, we further restrict our sample to stars with a metallicity range of --1.9 $\leq$ [Fe/H] $\leq$ --1.5. This is centered on the cluster's bulk population, at [Fe/H] = --1.7. This also helps prevent any metallicity-related bias, and leaves a total sample of 161 RGB and 38 eAGB stars.

Changing the metallicity limits we choose (--1.9 $\leq$ [Fe/H] $\leq$ --1.5) by $\pm$0.1 dex does not change our results by more than the random error. We also probed for variations in the mass differential within the sample, but could find no significant variation with metallicity or with any other elemental abundance (as listed in JP10). Seven of the stars in our final sample (LEIDs 23033, 24027, 38226, 42174, 43104, 43108 and 47151) have positive [Na/O] abundances suggesting they may belong to a second-generation, helium-rich population. These are all RGB stars, and excluding them makes no significant difference to the final result.

A systematic temperature difference of $\sim$130 K exists between the reddening-corrected photometric temperatures of M+09 and the ($V$--$K$)-based temperatures of JP10. This was traced to inaccuracies in the filter transmissions and zero point fluxes used for the optical photometry in M+09. The M+09 data were re-reduced, adopting a reddening value of $E(B-V)$ = 0.12 mag \citep{Harris96} and a distance of 5.3 kpc (see \S\ref{ErrorSect}, below), and replacing the original photographic $BV$ photometry of \citet{vLlPR+00} with the $UBVRI$ CCD photometry of \citet{BPB+09}. This reduced the temperature difference between the two datasets to an average of 47 K (r.m.s. 56 K). We retain the sample selection used above as the M+09 and Bellini et al.\ data are independent samples, and the M+09 data shows less scatter on the HRD.

\begin{figure}
\centerline{\includegraphics[height=0.5\textwidth,angle=-90]{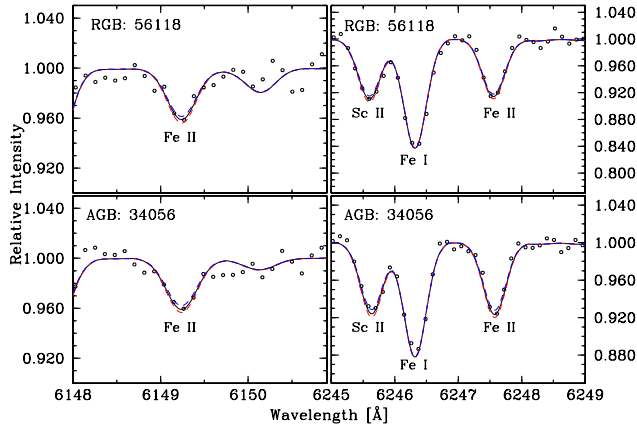}}
\caption{Example spectra of an RGB and AGB star with a typical temperature and metallicity  (LEID 56118, T$_{\rm eff}$ = 4620 K, [Fe/H] = --1.64; and LEID 34056, T$_{\rm eff}$ = 4875 K, [Fe/H] = --1.63, respectively). Synthetic spectral models of the 6149 and 6247 Fe {\sc ii} lines are overplotted with a fixed [Fe/H]{\sc ii} abundance. Open circles indicate the observed spectrum, the solid black line shows the best-fit log\,$g$ synthesis (derived from the EW analysis), and the dashed lines indicate changes in log\,$g$ of +0.1/--0.1 dex (blue/red lines).}
\label{SpecFig}
\end{figure}

The log\,$g$ values published in JP10 are based on photometric colours, which assume \emph{a priori} a mass of 0.8 M$_\odot$. As we measure mass, we require spectroscopic values of log\,$g$, which are free from such assumptions. These were determined by comparing the [Fe/H]$_{\rm I}$ and [Fe/H]$_{\rm II}$ abundances\footnote{[Fe/H] abundances derived from Fe {\sc i} and Fe {\sc ii} lines, respectively.} derived from the equivalent widths provided in JP10 (their Table 3). Model atmosphere grids were constructed for each star with log\,$g$  covering $\pm$0.5 dex (in 0.01 dex increments) from the photometrically-derived value provided in JP10 (their Table 2). Although the $T_{\rm eff}$ and microturbulence values were held fixed at the values given in JP10, we iteratively adjusted the model metallicity to equal the average value of the [Fe/H]$_{\rm I}$ and [Fe/H]$_{\rm II}$ ratios for a given log\,$g$ grid point. The final spectroscopic gravity was then selected as the log\,$g$ value that satisfied [Fe/H]$_{\rm I}$ = [Fe/H]$_{\rm II}$. The need for accurate log\,$g$ values meant that spectra were discarded where one or both of the Fe {\sc ii} lines could not be reliably measured and/or the two lines did not provide reasonable agreement on the ``best--fit'' log\,$g$ value. The comparative weakness of the Fe {\sc ii} lines (Figure \ref{SpecFig}) means that this represents a substantial reduction in our sample size. A further four RGB and two AGB stars were removed because they no longer fell into their respective regions in the HRD created using the Bellini et al.\ data. The final sample of high-quality targets comprises of 66 RGB and 21 eAGB stars. This subset of targets has negligible difference in their average metallicity, temperature and luminosity compared to our original selection. The new photometric and spectroscopic parameters for these objects are listed in Table \ref{ParamTable}, where stars are listed by their Leiden Identifier (LEID). These parameters were then used to measure a mass for each star, using Eq.\ (\ref{MEqn}).

\section{Error budget}
\label{ErrorSect}

\begin{figure}
\centerline{\includegraphics[height=0.5\textwidth,angle=-90]{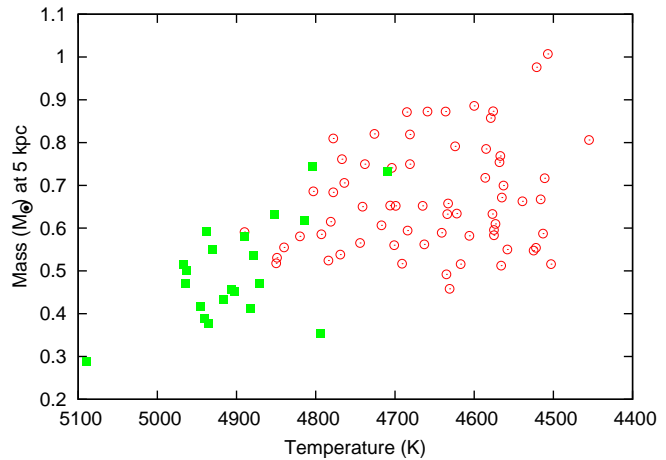}}
\caption{Variation of derived mass with temperature for our RGB (hollow red circles) and eAGB (filled blue squares) samples.}
\label{TempFig}
\end{figure}

The random error associated with individual RGB stellar masses is dominated by errors in log\,$g$ caused by uncertainties in the stellar metallicity, temperature and luminosity. The metallicity errors (taken from JP10) and a conservative microturbulence error of $\pm$0.3 km s$^{-1}$ were propagated through the log\,$g$ determinations described above. From these, a mass error is derived of $\delta M / M =\,^{+11}_{-10}\%$. Internal temperature errors of $\approx$56 K were estimated (being the r.m.s. deviation between values in Table \ref{ParamTable} and JP10), yielding an error of $\delta M / M =\,^{+13}_{-12}\%$. Finally, luminosity errors of $\sim$5\% yield a proportional effect on $\delta M / M$. Taking these errors in quadrature, we derive a r.m.s. random error of $\delta M / M =\,^{+18}_{-16}$\%. Errors on the AGB should be similar, though the log\,$g$ determination imparts a $^{+16}_{-15}$\% error in these stars, bringing the total error to $\delta M / M =\,^{+21}_{-20}$\%. This matches well with the r.m.s. scatter observed in Figure \ref{TempFig}: the r.m.s. scatter is 20\% for the RGB stars and 17\% for AGB stars. This shows that our random error budget covers the observed errors and suggests that the spread in the masses of stars on the eAGB is unlikely to be more than a few hundreds of a solar mass.

The systematic error budget is dominated by errors in temperature, reddening and distance to $\omega$ Cen. Systematic errors in temperature of $\sim$50 K were estimated, based on the 47 K systematic difference in the temperatures in Table \ref{ParamTable} and those of JP10, providing an error of $\delta M / M =\,^{+12}_{-11}\%$ as above. A conservative error of $\delta E(B-V) = \pm 0.025$ also yields a systematic temperature error of $\approx$50 K (M+09), providing an identical error. Distance estimates to $\omega$ Cen vary between 4.8 and 5.52 kpc \citep{vdVvdBVdZ06,DPPS+06}. We therefore assume a distance error of $\delta d / d =\,^{+4}_{-9}\%$ to encompass these values. This gives $\delta L / L = \delta M / M =\,^{+8}_{-18}\%$. Combining the temperature, reddening and distance errors in quadrature, we find a systematic error of $\delta M / M =\,^{+19}_{-24}\%$.

While difficult to quantify, additional sources of error may arise from the absolute accuracy of the {\sc marcs} model atmospheres used to measure the stellar luminosities and the wavelength dependence of the reddening correction. Choice of atmosphere models has very little effect: differences between masses derived using the Kurucz, Kurucz $\alpha$-enhanced and {\sc marcs} atmospheres are $<$1\%. Using ($V-K$) colours to derive a bolometric luminosity (\citealt{AAMR99}), however, yields masses which are $\sim$5\% greater than those derived from SED fitting. A significant difference here may be the effect of the H$^{-}$ opacity minimum at 1.6 $\mu$m. The observed flux in the 2MASS $H$ filter is $\approx$6\% less than expected (M+09). The accuracy of colour corrections to $U$ and $B$ band data in \citet{BPB+09} may also be important: Bellini et al.\ note that their $U$ magnitudes may be systematically incorrect by up to 0.15 mag. Removing the $U$-band data from our SED fits reduces the derived mass by $\approx$6\%. Similarly, altering the reddening law within a reasonable range can produce errors of $\sim$3\%. Combining the above uncertainties yields an approximate error of $\delta M / M \approx\,^{+6}_{-7}\%$, giving a final systematic error of $\delta M / M \approx\,^{+20}_{-25}\%$.

Our sample stars cover a relatively-narrow temperature regime, and the derived masses show no statistically-significant variation with temperature within the RGB population (Figure \ref{TempFig}). The eAGB population shows a weak temperature trend, though this is based on small-number statistics may be intrinsic: stars with lower envelope masses will have hotter temperatures, as on the HB. While the aforementioned systematic errors will affect the derived masses of our sample, the \emph{relative} masses of the RGB and AGB populations should remain largely unaffected. The ratio of the masses of the RGB and eAGB samples is therefore our most robust determination of mass loss between these two evolutionary points.

\section{The RGB vs.\ eAGB mass ratio}
\label{MassSect}



The average masses and their sample standard deviations found for our sample of RGB and eAGB stars, respectively, are 0.657 $\pm$ 0.090 M$_\odot$ and 0.488 $\pm$ 0.117 M$_\odot$. We stress that these values do not take into account the systematic errors of $\delta M / M =\,^{+20}_{-25}\%$ outlined above. The RGB mass is 18\% lower than the canonical $\approx$0.8 M$_\odot$ expected from stellar isochrone modelling (\citealt{GBB+02}; M+09), which is commensurate with these errors. It would be speculative to suggest why the masses we derive are systematically lower, though we note that the cluster's distance is (by a small margin) the largest contributor to the systematic error budget.

Taking the ratio of the two masses, which is largely independent of systematic errors, we find that 25.7 $\pm$ 4.3 \% (standard error) of mass is lost between the central RGB and eAGB. The true error may be a little larger (perhaps 1--2\%), due to temperature-dependent effects below the level of statistical detectability, and the robustness with which we can photometrically differentiate RGB and AGB stars.

\section{Discussion}
\label{DiscSect}

While the absolute masses we determine for our RGB and eAGB stars are systematically uncertain, the percentage difference between them can be put into context using well-calibrated initial and final stellar masses. This allows the mass of the HB stars to be determined and compared with HB models. The mass predicted by isochrone fitting for a 200 L$_\odot$ RGB star in $\omega$ Cen is between 0.80 and 0.85 M$_\odot$ (\citealt{GBB+02,DCJ+08}; M+09, and references therein), with very little mass loss occurring before this stage. This mass range agrees with the 0.806 $\pm$ 0.056 M$_\odot$ determined from the binary OGLEGC 17 \citep{TKB+01}. The final white dwarf mass in similar globular clusters is found to be 0.53 $\pm$ 0.03 M$_\odot$ (also derived using Eq. (\ref{MEqn}); \citealt{MKZ+04}).

Taking these two limits, this yields an eAGB star mass of 0.62 $\pm$ 0.04 M$_\odot$ and infers a mass loss of 0.21 $\pm$ 0.03 M$_\odot$ between the central RGB and eAGB (the difference in initial mass between the RGB and AGB stars is only a few $\times 10^{-3}$ M$_\odot$). The eAGB mass is commensurate with both the 0.61 and 0.63 M$_\odot$ derived using the Victoria-Regina and Dartmouth horizontal branch tracks, respectively (M+09)\footnote{These values assume helium of $Y \approx 0.24$, [$\alpha$/Fe] $\approx$ +0.3 and [Fe/H] $\approx$ --1.62.}, and within the spread of masses ($\pm$0.05 M$_\odot$) which these models predict.

The aforementioned isochrone fits provide the average RGB star we target with another 20 Myr on the RGB. Assuming mass loss on the HB itself is negligible, this translates to an average mass-loss rate of 10$^{-8}$ M$_\odot$ yr$^{-1}$ over the remainder of the RGB. In concurrence with previous relations (\citealt{Catelan09a}, and references therein), this is rather higher than the average Reimers' Law mass-loss rate \citep{Reimers75} of $1.9 \times 10^{-9}$ M$_\odot$ yr$^{-1}$ (assuming Reimers' $\eta = 0.5$). It is towards the higher end of the observed range of mass-loss rates estimated from chromospheric line profiles (e.g.\ \citealt{DSS09}; \citealt{VMC+11}), suggesting that stronger mass loss at the RGB tip is important.

One can also use the white dwarf mass to estimate that there is only 0.09 $\pm$ 0.05 M$_\odot$ of material for the star to lose during the $\sim$9 Myr it spends in its (post-)AGB phases: also yielding an average mass-loss rate of 10$^{-8}$ M$_\odot$ yr$^{-1}$. Chromospheric mass loss appears continuous on the AGB and can reach $\sim$6 $\times 10^{-8}$ M$_\odot$ yr$^{-1}$ \citep{DSS09}, while the cluster also has several dust-producing stars, which typically sustain $\sim$10$^{-6}$ M$_\odot$ yr$^{-1}$ for $\sim$10$^5$ yr (M+09; \citealt{MvLS+11}). The combination of these processes may mean that many stars lose their entire envelope before they reach the post-AGB phase, perhaps even becoming AGB-manqu\'e stars (`failed' AGB stars; e.g. \citealt{OConnell99}). This would explain the surprisingly-low luminosity (L $\sim$ 1500 L$_\odot$) of $\omega$ Cen's known post-AGB stars: LEID 16018 (Fehrenbach's star) and 32029 (V1) (\citealt{MvLS+11}).

Having measured the differential RGB/eAGB mass in $\omega$ Cen, and confirmed it against existing models, it now becomes possible to repeat the study in other clusters where the models are less certain. In particular, this technique will be useful in those clusters with higher metallicities and higher envelope masses, where the temperature of HB stars is less sensitive to stellar mass. Our current observations do not allow us to relate metallicity, initial mass, or initial abundance to RGB mass loss. For that, we need these observations repeated in other clusters. While other clusters are less populous than $\omega$ Cen, we are limited here by the existence of spectra, not by the number of stars. Bespoke targetting of stars (particularly on the eAGB), and a selection of spectral range with more gravity-sensitive lines, would allow this study to be repeated to higher accuracy, even in much-smaller clusters.

\section{Conclusions}
\label{ConcSect}

We have spectroscopically determined surface gravities for 66 central RGB and 21 eAGB stars in $\omega$ Cen, and combined these with photometric temperature and luminosity measurements to calculate the difference in mass between the two populations. We find that 26 $\pm$ 4\% of their mass is lost between these evolutionary phases, corresponding to 0.21 $\pm$ 0.03 M$_\odot$ for an initial mass of 0.83 M$_\odot$. By implication, this limits the mass lost on the AGB to some 0.09 $\pm$ 0.05 M$_\odot$, which may lead to early termination of the AGB and formation of AGB manqu\'e stars. Our derived HB masses of 0.62 $\pm$ 0.04 M$_\odot$ compares very well with HB models, which predict HB masses of 0.61 -- 0.63 M$_\odot$ for appropriate ($Y \approx 0.24$, [$\alpha$/Fe] $\approx$ +0.3, [Fe/H] $\approx$ --1.62) models. This method has the potential to provide physical constraints on currently-uncertain regimes in modelling horizontal branch stars.


\section*{Acknowledgements}

This material uses work supported by the National Science Foundation under award No. AST-1003201 to CIJ. We thank the referee and Olivia Jones for helpful comments.


\label{lastpage}

\end{document}